\documentclass[sigconf,nonacm]{acmart}
\preto{\abstractkeywords}{\nolinenumbers} 

\AtBeginDocument{%
  \providecommand\BibTeX{{%
    \normalfont B\kern-0.5em{\scshape i\kern-0.25em b}\kern-0.8em\TeX}}}

\usepackage{algorithmic}
\usepackage{algorithm}
\usepackage{multirow}
\usepackage{subfigure}
\usepackage{makecell}
\begin{document}


% \title{Scalable universal Variable Rate Lossless Image Compression via Masked Sampling}
\title{Deep Lossless Image Compression via Masked Sampling and Coarse-to-Fine Auto-Regression}
% \author{Tiantian Li}
% \affiliation{%
%   \institution{Hunan University}
%   % \streetaddress{1 Th{\o}rv{\"a}ld Circle}
%   \city{Changsha}
% \email{litiantian2er@hnu.edu.cn}

% \author{Name}
% \affiliation{%
%   \institution{Hunan University}
%   % \streetaddress{1 Th{\o}rv{\"a}ld Circle}
%   \city{City}
%   \country{Country}}
% \email{xx@xx.xx}

% \author{Name}
% \affiliation{%
%   \institution{Hunan University}
%   % \streetaddress{1 Th{\o}rv{\"a}ld Circle}
%   \city{City}
%   \country{Country}}
% \email{xx@xx.xx}
% \renewcommand{\shortauthors}{L.TianTian and Y.Gaobo, et al.}

\author{Tiantian Li}
\affiliation{%
  \institution{Hunan University}
  % \streetaddress{1 Th{\o}rv{\"a}ld Circle}
  \city{Changsha}
  \country{China}}
\email{litiantian2er@hnu.edu.cn}
% % \orcid{1234-5678-9012}
\author{Qunbing Xia}
\affiliation{%
  \institution{Aixiesheng Corporation}
  \city{Changsha}
  \country{China}
%   % \postcode{43017-6221}
}
\email{xiaqb@aixiesheng.com}
\author{Yue Li}
\affiliation{%
  \institution{University of South China}
  % \streetaddress{1 Th{\o}rv{\"a}ld Circle}
  \city{Hengyang}
  \country{China}}
\email{liyue@usc.edu.cn}
\author{RuiXiao Guo}
\affiliation{%
  \institution{Hunan University}
  % \streetaddress{1 Th{\o}rv{\"a}ld Circle}
  \city{Changsha}
  \country{China}}
\email{hnugrx@hnu.edu.cn}
\author{Gaobo Yang}
\affiliation{%
  \institution{Hunan University}
  % \streetaddress{1 Th{\o}rv{\"a}ld Circle}
  \city{Changsha}
  \country{China}}
\email{yanggaobo@hnu.edu.cn}
%% You do not have to enter your paper ID


\begin{abstract}
 Learning-based lossless image compression employs pixel-based or subimage-based auto-regression for probability estimation, which achieves desirable performances. However, the existing works only consider context dependencies in one direction, namely, those symbols that appear before the current symbol in raster order. We believe that the dependencies between the current and future symbols should be further considered. In this work, we propose a deep lossless image compression via masked sampling and coarse-to-fine auto-regression. It combines lossy reconstruction and progressive residual compression, which fuses contexts from various directions and is more consistent with human perception. Specifically,
 the residuals are decomposed via $T$ iterative masked sampling, and each sampling consists of three steps: 1) probability estimation, 2) mask computation, and 3) arithmetic coding. The iterative process progressively refines our prediction and gradually presents a real image. Extensive experimental results show that compared with the existing traditional and learned lossless compression, our method achieves comparable compression performance on extensive datasets with competitive coding speed and more flexibility. 
 \end{abstract}



\begin{teaserfigure}
  \centering
  \includegraphics[width=0.8\textwidth]{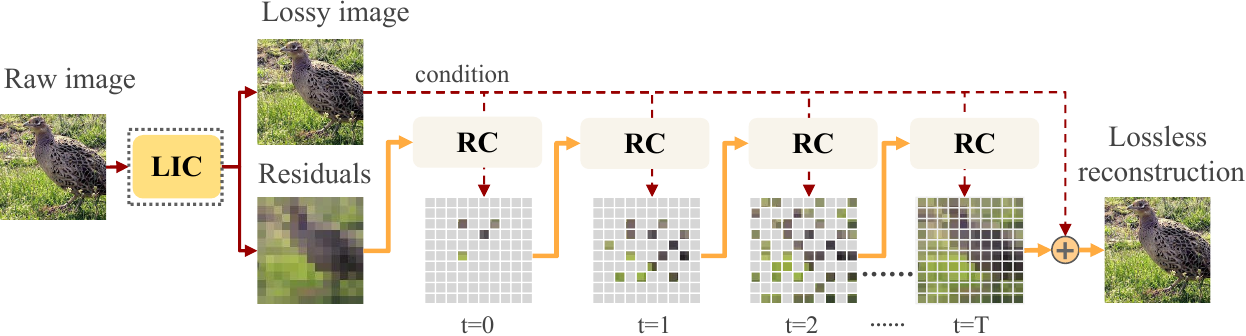}
  \caption{Overview of our masked sampling model. LIC is the learning-based lossy image compressor. RC is the proposed residual compressor. The gray parts are masked residuals during each sampling. For each iteration t, only unmasked residuals are encoded into bitstreams. After iterative $T$ times masked sampling, we finally get the whole residual image and add it to the lossy image to get the raw image.}
  \label{fig:overview}
\end{teaserfigure}
% \begin{figure}[ht]
%   \centering
%     % \fbox{\rule{0pt}{2.5in} \rule{0.6\linewidth}{0pt}}
%   \includegraphics[width=1\linewidth]{figures/overview-cropped.pdf}
%   \caption{Overview of our model. LIC is the learned-based lossy image compressor. RC is the proposed residual compressor. The gray parts are masked residuals during each sampling. For each iteration, only those unmasked residuals are encoded into bitstream. By iterative T times masked sampling, we finally get the residual, added the lossy image to get raw image.}
%   \label{fig:overview}
% \end{figure}

\maketitle

\section{Introduction}
For some application scenarios such as remote disease diagnosis and remote sensing, mathematically lossless image compression is essential for image data transmission and storage. In recent years, end-to-end deep lossy \cite{balle2018variational,he2022elic,cheng2020learnedloss}, and lossless \cite{l3c,iwave,cao2020lossless,hoogeboom2019integer,luo2023learned,RC,bai2022deepnearlyloss,bai2024deep,rhee2022lc} image compression have been witnessed with great progresses. They demonstrate superior compression performance to the conventional image codecs and great potential to achieve lower time complexity by efficient parallel computing. Following Shannon's source coding theorem \cite{shannon1948mathematical}, the low bound of expected code length is given by the real probability distribution of image data, i.e., the information entropy. Consequently, even the state-of-the-art lossless image compressors only have a 2x $\sim$ 3x compression rate, in contrast to 50x for lossy compression. 

The existing learning-based lossless image compression works \cite{salimans2017pixelcnn++, RC,bai2022deepnearlyloss,iwave} rely heavily on sophisticated auto-regressive models to estimate the probability distribution of pixels conditioned on decoded symbols, achieving remarkable compression with long decoding time. For practicality, recent works LC-FDNET\cite{luo2023learned} and DLPR \cite{bai2024deep} proposed some parallelization mechanisms, such as encoding in the unit of subimage or patch rather than individual pixels. However, regardless of pixel, patch, or subimage as the basic coding unit (CU), these works still strictly encode/decode an image in raster scan order, which is not consistent with the way humans perceive images. Moreover, previous works only consider one-directional contexts, ignoring other contextual dependencies in other directions. The mechanism is suitable for one-dimensional sequential data like texts, but not optimal or efficient for two-dimensional non-sequential images. Recent image diffusion models gradually denoise from a noisy image and finally obtain a clean image. Inspired by this, we treat image coding as a gradual process by masked sampling, which progressively refines the probability distribution of residuals for improved lossless compression. 
% Unlike machines that scan in raster order, Inspired by this, we view the decoding process as the gradual removal of the mask covering the image and eventually recovering the original image.
% In other words, we decompose the encoding/decoding into T iterations. Each iteration,  we sequentially perform three fixed operations: probability estimation, mask generation, and arithmetic coding. That is, at each iteration, we calculate a mask according to the estimated probability distribution, and use the mask to select partial samples from the whole image for coding, which we call this process as a sampling. During a sampling, the spatial locations of the final samples sampled are unknown and variable, depending on their probabilistic prediction scores.
% Moreover, previous works only consider one-directional contexts, ignoring other contextual dependencies in other directions. The mechanism is suitable for one-dimensional sequential data like texts, but not optimal or efficient for two-dimensional non-sequential images.
Furthermore, we propose a multi-direction auto-regression framework via masked sampling, which considers the dependencies among previous, current, and future symbols for better probability estimation of residuals in a coarse-to-fine manner. In summary, we follow the idea of deep lossy plus residuals coding for lossless image compression, but the residual coding is performed by iteratively refining the probability estimation. As shown in Figure~\ref{fig:overview}, a lossy image $\hat{X}$ is obtained by a lossy image compressor (LIC), and then the residual $R=X-\hat{X}$ is lossless encoded conditioned on $\hat{X}$ and the contexts (unmasked symbols) by the residual compressor (RC) with $T$ iterations. At each iteration $t$, we sequentially perform: $probability\ estimation\ (P)\to mask\ generation\ (M)\to arithmetic\ coding\ (E)$. Specifically, masks are generated according to the estimated probability distribution, which is used to select partial symbols from the whole residuals for coding. Thus, this process is called masked sampling. During each sampling, the spatial locations of the final symbols for coding are variable and scattered in multi-direction, %no longer one-direction, 
which makes the probability estimation for the next sampling be conditioned on more direction contexts. Only the symbols with high confidence in their probabilistic predictions are unmasked for the entropy coding. 
% Note that a forward propagation of neural network for predicting the probability of all symbols in parallel, but only those symbols that are confident in their probabilistic predictions are retained for entropy encoding. 
% For those unconfident symbols, we withdraw their probabilities and mask them out to avoid entropy coding with less accurate probability at this time. 
We can adjust the optimization granularity (fineness of the probability estimation) and speed (running time) of the model by controlling the variable $\textit{T}$. A bigger $\textit{T}$ implies a more refined iterative optimization process but a longer running time, and vice versa. As a result, $\textit{T}$ benefits compromising between compression rate and coding runtime. Moreover, the proposed approach is suitable for the lossless compression of grayscale and color images with a single model since it treats each component separately.
% It was verified that an end-to-end image compression model trained with a specific compression rate performs badly for other compression ratios. 
% To meet different bandwidth and storage requirements, multiple models with different compression ratios are required, which leads to model storage overhead and brings difficulty in controlling the bitrate accurately. In this work, due to the design of $\textit{T}$ times sampling, we can achieve variable bitrates through a single trained model, which is an extra advantage of our approach. Moreover, the proposed approach is suitable for the compression of grayscale and color images with a single model. 
In summary, the contributions of this work are three-fold.
\begin{itemize}
\item We propose a deep lossless image compression approach via iterative masked sampling and coarse-to-fine auto-regression. 
% The probability estimation is iteratively refined by 
Through $T$ times masked sampling, the autoregressive context modeling is extended from one direction to multiple directions and can be progressively conditioned on more contextual dependencies for enhanced performance. 

\item Scalable lossless image compression is achieved by adjusting $\textit{T}$, which controls the granularity and runtime of probability estimation of the residuals.
% The proposed approach is the first deep lossless image compression model with variable compression rate, and it is suitable for the compression of grayscale and color images with a single trained deep model. 

% \item The proposed approach is the first deep lossless image compression model with a variable compression rate, and it is suitable for the compression of grayscale and color images with a single trained deep model. 
\item The proposed approach is suitable for the compression of grayscale and color images with a single trained model. And it achieves competitive performance on various benchmark datasets. 
\end{itemize}

\section{RELATED WORKS}
\subsection{Hand-crafted Codec}
% Traditional image codecs like JPEG and JPEG2000 \cite{taubman2002jpeg2000} use mathematically well-defined transformations, such as Discrete cosine transform (DCT) and Discrete Wavelet Transform (DWT) \cite{DWT}, leading to
% the inclusion of various artifacts such as ringing and blocking.
Conventional image codecs exploit hand-crafted image statistics to remove spatial redundancies among neighboring pixels. Specifically, they achieve compression by reducing redundancies with reversible transform coding, prediction coding, and residual coding. The representative works include Discrete Wavelet Transform (DWT) \cite{DWT} based JPEG2000 \cite{taubman2002jpeg2000}, PNG \cite{boutell1997png}, BPG \cite{bellard2017bpg} and JPEG XL \cite{alakuijala2019jpeg}. These codecs heavily rely on hand-crafted linear transformations and have limitations when representing nonlinear correlations.
% The well-known traditional image codecs, such as Discrete Wavelet Transform (DWT) \cite{DWT} based JPEG2000 \cite{taubman2002jpeg2000}, BPG \cite{bellard2017bpg} and JPEG XL \cite{alakuijala2019jpeg}, heavily rely on hand-crafted linear transformations and have some inherent limitations when representing nonlinear correlations. 
\subsection{Learnable Lossy Compression}
 LIC has gained increasing interest due to its superior performance to conventional codecs. Taking advantage of sophisticated deep generative models, existing works \cite{balle2016end,balle2018variational,cheng2020learnedloss,he2022elic} train Variational AutoEncoder (VAE) to transform raw pixels into compact latent representations (the dimension is set to be much less than the number of pixels), and then model the probability distribution parameters of quantized representations via auto-regressive context models. The probabilities are calculated by a single or mixture distribution model and then are fed to an arithmetic coder to output bitstreams. Arithmetic coding is lossless, but the transform from pixels to latent representations by VAE and the quantization incur information loss.
 % the causes the information To compact representations, VAE outputs latent representations with the dimension less than the number of pixels, achieving compact representations. Such conversion from pixels to features incurs information loss, which implies that the existing work cannot deal with lossless image compression. A more severe limitation of the compressive auto- 
 % For example, an end-to-end trainable model is presented for image compression based on VAE \cite{balle2018variational}, which incorporates a hyperprior to effectively capture spatial dependencies in the latent representation. 
 Benefiting from the likelihood-based generative model's ability to learn the unknown probability distribution of given data, the learning-based lossy image compression achieves high fidelity and a desirable compression ratio. In this work, we integrate a sophisticated lossy image compression, followed by a customized residual coding for lossless image compression.

\subsection{Learnable Lossless Compression}
The conventional lossless compression of given symbols can usually be solved in two
stages: 1) statistical model the probability distribution $\sim p$ of symbols; 2) according to the statistical distribution $\sim p$, encode the symbols into bitstreams with arithmetic coder. Deep generative models are mainly trained to complete the first-stage task.
% the statistical modeling of given image data. 
Among them, the pixel-wise autoregressive model, i.e, PixelCNN++ \cite{salimans2017pixelcnn++} models the distribution as $p(x)=\prod _ip(x_i|x_1,\dots,x_{i-1})$. Despite the notable compression rate, its sequential decoding process makes it time-consuming and impractical. The subimage-wise models address this issue by modeling the distribution in a coarse granularity as $p(S_i)=p(S_i|S_1,\dots, S_{i-1})$. L3C \cite{l3c} and SRec \cite{cao2020lossless} use convolution layers and average pooling to decompose the image $X$ into subimages ${\{S_1, S_2,\dots S_n\}}$ of multiple scales, respectively. LC-FDNet \cite{rhee2022lc} performs spatial downsampling (\textit{space-to-depth}) on image $X$ to obtain subimages with the same scale. However, their compression ratios are inferior to pixel-wise models. iWave++ \cite{iwave} exploits a multi-level lifting scheme \cite{sweldens1996lifting} to learn reversible integer DWT and models the probability of each wavelet subband via PixelCNN++ \cite{salimans2017pixelcnn++}. But subbands are serially decoded and each wavelet coefficient in subband is pixel-by-pixel, making it time-consuming. Joint lossy compression and residual coding models \cite{RC}, Near-lossless \cite{bai2022deepnearlyloss} and its extension work DLPR \cite{bai2024deep} derive the probability distribution of an image $X$ conditioned on the lossy reconstruction $\hat{X}$ from traditional codec and learned codec, respectively. In this paper, we also join lossy compression with a few bits and then design a novel residual coding framework to encode the remaining residuals in a coarse-to-fine manner.
% \subsection{Variable-rate compression}
% There exists a handful of image compression approaches to support variable rates using a single model. However, they are mainly focusing on LIC. Rate adaptation in LIC is an important technology, and the use of a single model to meet various compression ratios is one of the decisive factors for the widespread use of LIC, especially in resource-constrained scenarios. In some bandwidth-limited scenarios, the rate control strategy plays an important role in image and video transmission. The aim of rate control is to achieve the best reconstruction quality under the given rate constraint. The $R-\lambda $ model is a practical method for LIC to control the rate by giving the Lagrange multiplier $\lambda $. The rate-distortion optimization in learned lossy image compression methods is realized by minimizing a combined loss function based on compression ratio and distortion between an original and a corresponding output image. For lossless compression, since there is no distortion of information, variable rate by rate-distortion optimization is unworkable, causing the constant compression rate for lossless compression. To our knowledge,  our model is the first to implement variable rate lossless compression. And our networks only needs to be trained once (not per image), regardless of input image dimensions and the desired compression rate.

\section{Methodology}
\subsection{Overview}
\begin{figure*}[h]
  \centering
    % \fbox{\rule{0pt}{2.5in} \rule{0.9\linewidth}{0pt}}
  \includegraphics[width=0.85\linewidth]{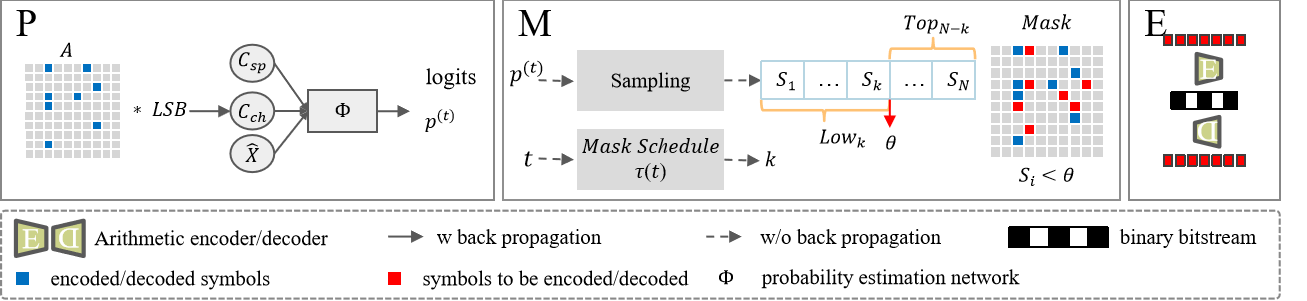}
  \caption{Main components of the proposed framework RC via \textit{$T$} times masked sampling.}
  \label{fig:sampling}
\end{figure*}

\begin{figure}[h]
  \centering
    % \fbox{\rule{0pt}{2.5in} \rule{0.9\linewidth}{0pt}}
  \includegraphics[width=0.85\linewidth]{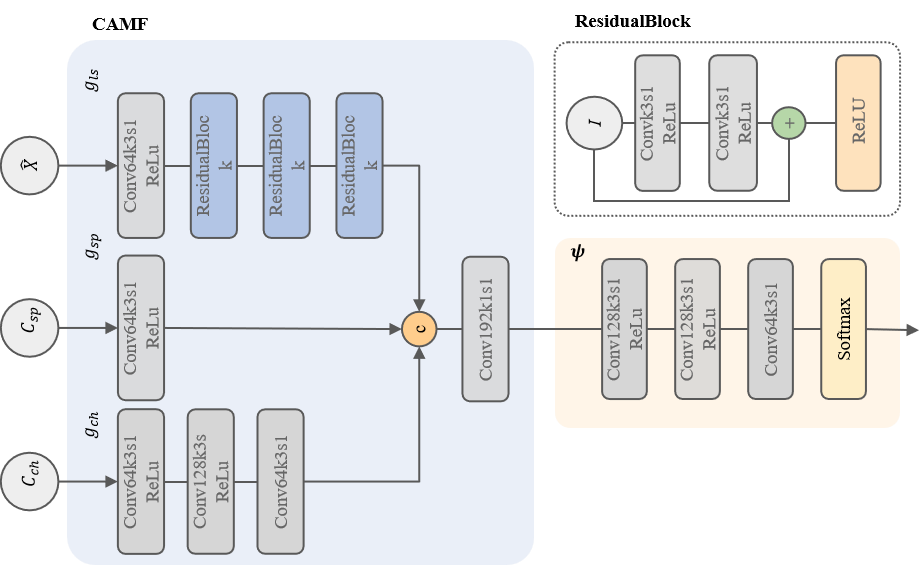}
  \caption{Detailed architecture of the residual probability estimation network $\Phi$ .}
  \label{fig:rc}
\end{figure}

We present a lossless image compression framework by integrating lossy image compression with residual coding. As shown in Figure~\ref{fig:overview}, we first obtain a lossy image $\hat{X}$ with the existing lossy image compressor ELIC \cite{he2022elic}. Secondly, we construct a residual coding system that progressively encodes/decodes residuals by iterative masked sampling. Specifically, we decompose  9-bit residual into 5-bit plane $R_{1:5}$ (MSB) and 4-bit plane $R_{6:9}$ (LSB). In addition, there is a 1-bit flag to signal whether MSB should be encoded. Only when the flag is 1, we need to encode MSB by Run-length Coding (RLE) \cite{RLE}. Next, we encode the LSB  with  $\textit{T}$ times masked sampling. At each iteration $t$, we cascade 3 operations: $P\to M \to E$, as shown in Figure~\ref{fig:rc}. The $P$ operator sends the lossy image $\hat{X}$ condition, channel condition $C_{ch}$ and space condition $C_{sp}$ into the proposed probability
estimation network $\Phi$ to output $p^{(t)}$, the logits of symbols over all possible values in time $t$. The $M$ operator receives $p^{(t)}$ to generate a mask matrix $M \in R^{H\times w}$, which is used to select the symbols to be encoded (\textit{the red points of Mask in Figure~\ref{fig:sampling}}) in time $t$, which are subsequently encoded into the bitstream by the operator $E$. After \textit{$T$} iterations, \textit{$R$} is fully encoded.

Existing lossless image compression works mainly use integrated color components as input and are incapable of supporting grayscale images with a single model. In contrast, our model is more flexible and compatible with grayscale images and color images with a single-trained model. To support multiple format inputs, we separately process each image component with the above pipeline. To accelerate coding, we run each component's network in parallel during inference.

\subsection{Lossy Compression}
Our compression framework exhibits compatibility with arbitrary lossy image codecs. For experimental validation, we adopt the well-established end-to-end learned image compressor ELIC \cite{he2022elic}, which synergistically combines a hyperprior model with a context model to achieve state-of-the-art rate-distortion performance.
% which is based on the framework of hyperprior model and  context model.
% with efficient hyper-prior framework. to perform the part of lossy compression.  

% It is based on the framework of the hyper-prior model and existing context model. Motivated by the observation of energy compaction in learned image compression, the authors unevenly group the representations in latent space along the channel and propose a multi-dimension entropy estimation model named Space-Channel ConTeXt (SCCTX) model. 
% ELIC is fast and effective in reducing the bitrate. More details refer to the original text.

% ed space-channel contextual adaptive coding.
% uneven channel-condition adaptive coding. 
% context model 
\subsection{Residual Compression}
\label{residual comp}
Since the superiority of LIC, the theoretical interval [-255, 255] of $R=X-\hat{X}$ can hardly be filled up in practice. We linearly transform and decompose the residuals into high 5-bit
% new range. We first subtract an offset $R_{min}$, the min value of $R$, and move the residual to the positive axis and the range of [0,510]. Then, we slice the 9-bit residual into the high 5-bit (MSB) 
\begin{equation}
    MSB=\left \lfloor \frac{R-R_{min}}{2^4}  \right \rfloor 
\end{equation}
and the low 4-bit plane (LSB)
\begin{equation}
   LSB= (R-R_{min})\bmod\ {2^4}.
\end{equation}
In particular, when the offset residual ${R-R_{min}}$ completely falls in the range of [0,63], MSB is all 0, meaning there is no information to be compressed. Therefore, setting a 1-bit $flag$ can greatly simplify computation and storage. MSB's sparsity arising from the narrow range of residuals makes it unsuitable and uneconomical for modeling through neural networks. Hence, we choose simple and straightforward RLE \cite{RLE} to encode MSB when $flag=1$.

LSB mainly contains visually important data and contributes to finer textual details. Consequently, we compress it with the designed neural model RC, comprising three operations as shown in Figure~\ref{fig:rc}. The pipeline for coding the LSB is summarized as the following three steps.

\noindent\textbf{Softmax-likelihood Probability estimation.} 
We employ the flexible softmax likelihood-based non-parametric probability model $\Phi$, as shown in Figure~\ref{fig:rc}. $\Phi$ is composed of the backbone Context-aware Multi-domain Fusion Module (CAMF) and the probability predictor $\psi $. A forward of $\Phi$ outputs
% a probability distribution of $R$. Specially, we parallelly input the lossy image $\hat{X}$, channel context and space context into the CAMF, to extract and fuse the three different-domain contextual features. The aggregated features are subsequently fed to the probability predictor $\psi $, coupled with a softmax layer, to directly derive 
the probability mass function (PMF) of whole LSB of $R$, namely logits $p^{(t)}\in {R^{ H\times W\times 64}}$ on possible value for each symbol in time $t$. The derived logits share height and width with $R$, and the channels are 64.
% The residual probability estimation network $\Phi$ employs a Context-Aware Multi-Fusion (CAMF) backbone that 
CAMF integrates information from three pipelines: lossy reconstruction prior $\hat{X}$, spatial context $C_{sp}$, and channel context $C_{ch}$. In previous works, a spatial context often refers to the currently observable neighbors of $i-th$ coding symbols. It is limited as symbols before $r_i$ in a predefined raster scan order (i.e., from left to right line by line), namely $r_{<i}=\{r_1,\dots,r_{i-1}\}$. Our method expands them to symbols coming from all directions, 

\begin{equation}
  C_{sp}=\{r_j | j\in N, A_j=1\} ,
\end{equation}
where $A$ is the binary matrix plotted in Figure~\ref{fig} and $A_j=1$ indicates the encoded/decoded symbols (blue points).

% used to indicate the location of encoded/decoded symbols and is updated after finishing the E at each iteration. $N$ is the number of residual r in the whole $R$. 
The spatial context $C_{sp}$ is perceived by the spatial context model $g _{sp}$, and rich neighborhood information $F_{sp}$ is extracted to promote the PMF estimation of symbols.

\begin{equation}
     F_{sp}=g_{sp}(C_{sp}).
\end{equation}
% where A is the accumulated binary anchor matrix, whose symbols at the locations with value 1 are visible at the current time.

Channel context $C_{ch}$ refers to the observable neighboring components. 
% For luminance component (Y), it lacks referable components, hence we pad $C_{ch}$ with the space context. 
For chrominance components, it leverages cross-channel correlations:
% it refers to the previously decoded components. The more channel contextual dependencies can be exploited by the channel context model, which are denoted as
% can exploit the channel contextual dependencies by
\begin{equation}
     F_{ch}=g_{ch}(C_{ch}).
\end{equation} 

Lossy reconstruction prior $ F_{ls}$ is extracted from the lossy image $\hat{X}$ by the lossy context model $ g_{ls}$, composed of continuously stacked residual blocks for increasing the receptive field.
\begin{equation}
     F_{ls}=g_{ls}(\hat{X}).
\end{equation} 
% Specially, we propose a backbone Context-aware Cross-level Fusion Module (CACF) to extract and integrate the multi-level features. In parallel, we input the lossy image $\hat{X}$, channel context, and space context to three individual context models $g_{ls},g_{ch},g_{sp}$ to capture different-domain contextual dependencies. Then, we concatenate these contextual dependencies and fed them to the $1\times 1$ convolution to fuse these multi-domain feature representations, which collectively adjusts the number of channels to save the calculation. After thorough CAMF, the aggregated features are subsequently fed to the probability predictor $\psi $, coupled with a softmax layer, to derive the probability mass func-
% tion (PMF) of whole $R$, logits $p^{R\in H\times W\times 64}$ on possible value for each symbol. The derived logits share height and width with $R$, and the channels are 64.
The fusion module combines these contexts via $1\times1$ convolution followed by softmax-based probability prediction:

\begin{equation}
p^{(t)} = \text{Softmax}\left(\psi\left(\text{Conv}{1\times1}(F_{sp}, F_{{ch}}, F_{{ls}})\right)\right)
\end{equation}
where $\psi$ implements lightweight (64-channel) convolution for parameter efficiency. It directly outputs 64 numbers per symbol describing the logits on the 64 potential values for each symbol.

% Next, we concatenate the three context representations $\Phi$  and further fuse them by the 1x1 convolution, which collectively adjusts the number of channels to save the calculation. The aggregated features are then fed to the probability predictor $\psi$ to generate $p^{(t)}$, the distribution of symbols. 

% As aforementioned in Section~\ref{architexture}, existing methods usually employ parametric models as probability predictor $\psi$, generating the entropy parameters of the assumed distribution, such as logistic mixture likelihood \cite{l3c, RC,hoogeboom2019integer}, gaussian mixture likelihood \cite{cheng2020learned,iwave} and so on. Differently, our model uses softmax likelihood rather than discrete mixture distribution. We configure a 64-channel convolution and softmax layer as the head of the probabilistic predictor $\psi$. It directly outputs 64 numbers per symbol
% \begin{equation}
%     p=Softmax(\psi  (Conv1x1 (\Phi_{sp},\Phi_{ch},\Phi_{ls})),
% \end{equation} 
% describing the logits on the 64 potential values for each symbol. Due to the moderate number of convolution kernels, this configuration does not significantly increase the number of parameters and slow down the training.
% For parametric models, they are usually optimized by bitrate loss $ L=E[-\log p]$.
% % \begin{equation}
% %     L=E[-\log p]
% % \end{equation} 

\noindent\textbf{Mask generator.}
At time $t$, we only encode $k$ symbols with high-confidence probability prediction.
% we first compute $k=\tau(t)*N$, the number of symbols to be masked according to the mask scheduling function $\tau(t)$ and the total number of pixels $N=H\times W$. 
We sample a value matrix $V^{(t)}\in R^{H\times W}$ by the probability $p^{(t)}$ predicted by $\Phi$ and calculate a confidence score matrix $S^{(t)}\in R^{H\times W}$, which indicates the confidence of probability prediction in each spatial location (i,j). The $k$ locations with lower scores ($S^{(t)}_{i,j}\le \theta$) are masked ($M^{(t)}_{i,j}=1$). The $N-k$ locations with $S^{(t)}_{i,j}>\theta$ (the blue and red point in Mask in the Figure~\ref{fig:rc}) are reserved ($M^{(t)}_{i,j}=0$).

The training objective of $\Phi$ minimizes masked cross-entropy
% For training, we use cross-entropy loss to measure the distance 
of predicted distribution $p$ and the marginal distribution $\tilde{p}$ of symbols to be encoded:
% With the designed mask, our model is finally trained with the following loss function 
\begin{equation}
\begin{split}
     L_{M^{(t)}_{i,j}=1}=
     % H(\sim{p},p)\\=E_{r_j\sim\tilde{p}[-logp(r_i)]}\\=
     \sum_{r_{i,j}\in R}\tilde{p}(r_{i,j})\log p^{(t)}(r_{i,j}).          
\end{split}
\end{equation} 

Note that during training, the model only performs the single-pass mask implementation that enables efficient training without iterative refinement.
% one sampling with no iteration. Therefore, the mask matrix $M$ has no superscript $(t)$. We will detailedly explain in Section~\ref{c-f}.

\noindent\textbf{Arithmetic coding.} After operator $M$, we obtain the locations of symbols with high confidence scores. We maintain an anchor matrix $A$ to record the location of encoded/decoded symbols (blue points), which is updated cumulatively in each sampling. We use torchac\cite{l3c} as an arithmetic coder to encode/decode the red points.

% After getting the logits, we 

\subsection{Iterative masked sampling}
Motivated by the MaskGIT \cite{chang2022maskgit} that iteratively generates the latent tokens by a mask schedule, we adapt iterative masked sampling to residual compression in the pixel domain. As described in Algorithm~\ref{alg:encode}, at initilization ($t=1$), we initialize the $A=0$ and $M^(1)=1$, representing no known symbols for context.
At iteration $t$, we first compute $k = \tau(t)\times H \times W$, the number of symbols to be masked according to the mask scheduler
$\tau(t)$ and the total number of pixels $H \times W$ .Secondly, we sample $V^{(t)}$ by probability $p^{(t)}$ to replace the unknown symbols ($A_{i,j}=0$) with $V^{(t)}_{i,j}$. We get the probability $p^{(t)}_{i,j, V^{(t)}_{i,j}}$, a scalar representing the probability of the location (i,j) valued $V^{(t)}_{i,j}$, which is then used to compute a score matrix $S^{(t)}$ by 

\begin{equation}
S^{(t)}_{i,j}=\left\{\begin{matrix}
 p^{(t)}_{i,j,V^{(t)}_{i,j}}+\beta* z, z\sim N(0,1)  &if A_{i,j}=0 \\
 1 &if A_{i,j}=1,
\end{matrix}\right. 
\label{eq:scores}
\end{equation}
where $\beta$ is a temperature factor to regulate the degree of added random noise $z$. The $k$ lowest-score positions are masked ($M^{(t)}_{(i,j)}$) for subsequent refinement. The decoding is described in Algorithm~\ref{alg:decode}. The $\textit{T}$-step iterative encoding/decoding process 
% enables dynamic error correction - later iterations can override earlier predictions using enriched contextual information. As shown in Algorithm~\ref{alg:decode}, the $\textit{T}$-step decoding 
progressively converts coarse initial estimates into fine-grained reconstructions.
% An encoding/decoding goes through $\textit{T}$ times masked sampling. The complete decoding algorithm is described in Algorithm~\ref{alg:decode} respectively. With the proposed design of iterative masked sampling, we make our probability estimation of residual modifiable and revocable. 

\subsection{Coarse to Fine and Multi-directional Autoregression}
\label{c-f}
Following the raster scan ordering, pixel/subimage-based autoregressive models sequentially decode an image with slow speeds. Usually, these autoregressive models employ $k \times k$ masked convolution to mask out further information, ensuring strictly causal dependency - the conditional probability estimation of current symbols only depends on decoded symbols. In this paper, we propose a coarse-to-fine and multi-directional auto-regressive context model.
The mask scheduling function $\tau(t)$
controls the coarse-to-fine refinement process through following critical properties. $\tau(1)=1$ initiates full masking (coarse estimation) and $\tau(T)=0 $ ensures complete refinement (fine details). $\tau(t)\downarrow $ as $t$ increases maintains progressive information accumulation, where probability prediction is more confident and fine-grained because of incorporating more contextual
information. We finally take the monotonically decreasing and positive part of the cosine function:
% as our mask schedule function $\tau(t) $ to generate the k, which satisfies the constraint: $\tau(1)=N , \tau(T))=0$ and $\tau(t)\downarrow $.
\begin{equation}
\tau(t) =\left\{\begin{matrix}
cos(\frac{t\pi}{2T})  & if\enspace inference\\
  cos(\frac{\epsilon  \pi}{2})& if\enspace training
\end{matrix}\right.
\label{schedule}
\end{equation}
where $\epsilon \in (0,1)$ is a random number. 

% Note that there are some differences between training and inference. 
When training, we only perform one masked sampling for each component. The $k$ is decided by the random ratio $\epsilon$. We generate $k$ positions (i,j) with the lowest values from a matrix obeying a uniform distribution $U\sim (0,1)$ and masks them $M_{i,j}=1$. When encoding/decoding, we iteratively sample $\textit{T}$ times, and the $k$ and $M^{(t)}$ are calculated as Algorithm~\ref{alg:encode} and Algorithm~\ref{alg:decode}.
% At the beginning ($t=1$), our predicted probabilities are coarse-grained because there are fewer known symbols to refer to due to the highest $k$. As t increases, the $k$ is gradually reduced until it reaches 0. Consequently, our probability prediction is more confident and fine-grained because of incorporating more contextual information. 
% From a temporal perspective, we formalize this as
% \begin{equation}
% p^{(t)}(x)= p(x_{i,j}|p^{(t-1)},\dots,p^{(1)},p^{(0)})
% \label{eq:temporal}
% \end{equation} 
At each time $t$, the probability modeling of the unknown symbols is conditioned on the encoded/decoded symbols in previous times $<t$,
\begin{equation}
p^{(t)}_{{A_{i,j}=0}}(x)= p^{(t)}(x_{i,j}|x_{m,n},A_{m,n}= 1)
\label{auto-re}
\end{equation} 
The known symbols from multiple directions facilitate effectively learning the pattern of residual by $\textit{T}$ iterations.
% \subsection{Iterative Sampling}
% Note that there are some differences between training and inference.
% % that the process of training and inference is different. 
% When training, we perform one sampling for each component. The k is controlled by the random ratio $\epsilon$. The mask is decided by a matrix $U$ with a uniform distribution of 0-1. The $top_{N-k}$ locations are set $M=0$, other k locations are  $M=1$. When encoding/decoding, we iteratively sample T times, as described in Algorithm~\ref{alg:encode} and Algorithm~\ref{alg:decode}.
% Each sampling process consists of three basic operations. 

\begin{algorithm}[!h]
    \caption{algorithm of encoding}
    \label{alg:encode}
    \renewcommand{\algorithmicrequire}{\textbf{Input:}}
    \renewcommand{\algorithmicensure}{\textbf{Output:}}
    \begin{algorithmic}[1]
        \REQUIRE $T$, $H, W$, $\hat{X}$,$\beta $,$R$\\
        $\quad A\gets \textbf{0}$, $M^{(1)}\gets \textbf{1}$\\
        $\quad$Set the random number seed; %%input
        \ENSURE $bitstream $
        \FOR{$t=1$ to $T$}
            \STATE  $k=\tau(t)\times H\times W$  
            \STATE $C_{sp}=A\cdot R$
            \IF{$C_{ch} =\emptyset$ }
                \STATE $C_{ch} =C_{sp}$
            \ENDIF
            \STATE  $p^{(t)}=Softmax(\psi(g_{ls}(\hat{X}),g_{sp}(C_{sp}),g_{ch}(C_{ch}))$
            \STATE $ V^{(t)} \leftarrow$ probabilistic sampling from $ by \sim p^{(t)}$

            \IF {$A_{i,j}=0$}
                \STATE $\tilde{p}^{(t)}_{i,j}=inf$
            \ELSE
                \STATE$\tilde{p}^{(t)}_{i,j}=p^{(t)}(i,j,V^{(t)}_{i,j})$
            \ENDIF
            \STATE $S^{(t)}_{i,j}=\log {\tilde{p}^{(t)}(i,j)}+\beta *z,z\sim N(0,1)$
            \STATE Flatten  $S^{(t)}$ and sort it in ascending order
            \STATE$\theta =S^{(t)}[k]$
            \IF {$S^{(t)}_{i,j}<\theta$}
                \STATE $M^{(t)}_{i,j}=1$
            \ELSE
                \STATE$M^{(t)}_{i,j}=0$
            \ENDIF
            \STATE $symbol=(A \oplus   M^{(t)})\cdot R, pdf=(A\oplus M^{(t)})\cdot p^{(t)}$
            \STATE$  bitstream +=E(symbol, pdf)$
            \STATE $A=A || M^{(t)}$
           
       \ENDFOR
        \RETURN  $bitstream $
    \end{algorithmic}
\end{algorithm}

\begin{algorithm}[!h]
    \caption{algorithm of decoding}
    \label{alg:decode}
    \renewcommand{\algorithmicrequire}{\textbf{Input:}}
    \renewcommand{\algorithmicensure}{\textbf{Output:}}
    \begin{algorithmic}[1]
        \REQUIRE $bitstream, T$, $H, W$, $\hat{X}$,$\beta $,$\quad A\gets \textbf{0}$, $M^{(1)}\gets \textbf{1}, R\gets \textbf{0}$\\ %%input
        $\quad$Set the random number seed. %%input
        \ENSURE $R$ %$A, p$    %%output
        \FOR{$t=1$ to $T$}
            \STATE  $k=\tau(t)\times H\times W$, $C_{sp}=A\cdot R$
            
            \IF{$C_{ch} =\emptyset$ }
                \STATE $C_{ch} =C_{sp}$
            \ENDIF
            \STATE  $p^{(t)}=Softmax(\psi(g_{ls}(\hat{X}),g_{sp}(C_{sp}),g_{ch}(C_{ch}))$
            \STATE $ V^{(t)} \leftarrow$ probabilistic sampling by$ \sim p^{(t)}$
            \STATE calculate $S^{(t)}_{i,j}$ by Equation~\ref{eq:scores}
            % \IF {$A_{i,j}=1$}
            %     \STATE $\tilde{p}^{(t)}_{i,j}=inf$
            % \ELSE
            %     \STATE$\tilde{p}^{(t)}_{i,j}=p^{(t)}(i,j,V^{(t)}_{i,j})$
            % \ENDIF
            % \STATE $S^{(t)}_{i,j}=\log {\tilde{p}^{(t)}(i,j)}+\beta *z,z\sim N(0,1)$
            \STATE Flatten  $S^{(t)}$ and sort it in ascending order, $\theta =S^{(t)}[k]$
            \IF {$S^{(t)}_{i,j}<\theta$}
                \STATE $M^{(t)}_{i,j}=1$
            \ELSE
                \STATE$M^{(t)}_{i,j}=0$
            \ENDIF
            % \STATE $M^{(t)}=S^{(t)}<=\theta$
            \STATE $ pdf=(A\oplus M^{(t)})\cdot p^{(t)}$
            \STATE$  R_{(A \oplus   M^{(t)})=1}=D(bitstream, pdf)$
            % \STATE $R_{(A \oplus   M^{(t)})=1}=symbol$
            \STATE $A=A || M^{(t)}$
           
       \ENDFOR
       % \STATE $X=\hat{X}+R$
        \RETURN  $R $
    \end{algorithmic}
\end{algorithm}
% To be specific,  Our mask schedule function takes the monotonically decreasing and positive part of the cosine function. At the beginning of sampling, our predicted probabilities are coarse-grained because there is less contextual information to refer to (The number of symbols to be masked off k is the highest). As time step t increases, our probability prediction during each sampling becomes more fine-grained.
\subsection{Rate-Latency Scalability}%Variable bitrate and latency}
\begin{figure}[ht]
    \centering
        \includegraphics[width=0.75\columnwidth]{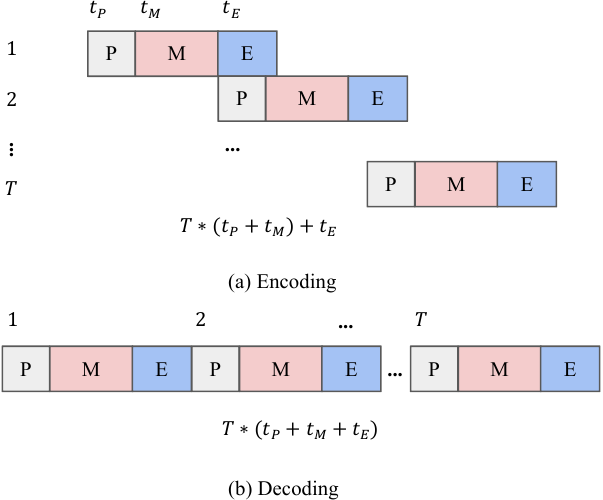}
    \caption{Latency analysis}
    \label{fig:time}
\end{figure}

\label{iterative}
Differing from existing approaches in computational scalability, our framework introduces tunable rate-latency tradeoffs through adaptive iteration control. 
DLPR \cite{bai2024deep} splits the $H\times W$ image into multiple non-overlapping $P \times P$ patches and codes all patches in parallel. Since the context of each pixel is limited to a patch, the $H\times W$ times sequential context computations can be finished with $P^2$ times forward computation latency, a significant reduction compared with $H\times W$ times runtime of PixelCnn++ \cite{salimans2017pixelcnn++}. In contrast, our method only costs $T\in \{1,2,\dots,12\}$ times forward computation latency. $P^2$($P=64$ in DLPR) is usually greater than $T$. During each sampling, the probability of all symbols in $R$ is generated simultaneously in parallel. As shown in Figure~\ref{fig:time}, the time overhead of the three operators $ P, M, E$ in each sampling are denoted as $t_P,t_M,t_e$. 
% Since it runs on the CPU and needs to read/write the stream, the $t_M>t_E>t_P>t_M$
% operator E takes the longest runtime. Without the forward propagation of the neural network, the operator M has the lowest latency ( $t_E>t_P>t_M$). 
When encoding, we minimize runtime by cascading three operators. Specifically, once the $M$ operator is executed at time $t-1$, the $P$ operator at time $t$ can commence without waiting for the completion of the $E$ operator at $t-1$.
% Sfor minimizing runtime, we cascade the three operators in a pipelined fashion. That is, once the M operator in the time $t-1$ has been executed, the P operator in time $t$ can begin without waiting until the E operator in $t-1$ has been finished. Consequently
As a result, we get a latency of $T*(t_P+t_M)+t_E$. When decoding, the probability estimation operator $P$ at time $t+1$ can only be operated after the arithmetic decoding operator $E$ at time $t$ is completed. Consequently, decoding requires strict serialization with a latency of $T*(t_P+t_M+t_E)$. 
In addition, the iteration count $\textit{T}$ controls the rate-latency tradeoffs. A larger $\textit{T}$ corresponds to finer coding, lower bit rates, and increased latency, while a smaller $\textit{T}$ leads to coarser coding, higher bit rates, and decreased latency, and vice versa. 
% The larger the $\textit{T}$, the finer the coding, and the lower the bit rate. 
% In summary, the total latency increases monotonically with the variable $\textit{T}$. Conversely, the total bitrate exhibits a monotonically decreasing nonlinear relationship of $\textit{T}$.
%, the lower bit rate, but the delay will increase. The smaller $\textit{T}$ is, the coarser the coding process, the bit rate increases but the delay decreases. 
We can flexibly trade off the rate and latency of compression by fine-tuning an appropriate $\textit{T}$
% This flexibility allows us to prioritize either aspect according to our requirements by simply tuning $T$
, which is a significant advantage of our method over other lossless compression methods. 
% For DPLR \cite{bai2024deep}, the bitrate is a constant once the model is fixed, failing to support variable rate to adapt to different bandwidth scenarios with a single trained model. And our method can do that.

% \begin{figure}[ht]
%     \centering
%     \begin{subfigure}{\columnwidth}
%        \includegraphics[width=0.8\columnwidth]{figures/encode-cropped.pdf}
%         \caption{Encoding}
%     \end{subfigure}
%     \hfill
%     \begin{subfigure}{\columnwidth}
%         \includegraphics[width=0.88\columnwidth]{figures/decode-cropped.pdf}
%         \caption{Decoding}
%     \end{subfigure}
%     \caption{Latency analysis}
%     \label{fig:time}
% \end{figure}

\section{Experiments}
\begin{table*}[htbp]
\caption{Lossless image compression performance (bpp) of our proposed coding system, compared
with other lossless image codecs on all evaluation sets. The bpp of our method in the table contains the sum of all parts. The best performance is highlighted in \textbf{bold}, and the second-best performance is denoted with *. ``-'' indicates that the result is not available because of the availability of code or training weights }
\label{tab:bpp}
\resizebox{1.2\columnwidth}{!}{
\begin{tabular}{ccccccc}
\toprule
\textbf{Codec} & \textbf{ImageNet64} & \textbf{Open Images}                   & \textbf{CLIC.m} & \textbf{CLIC.p} & \textbf{DIV2K} & \textbf{Kodak}                        \\
\midrule
PNG \cite{boutell1997png}           & 17.22               & 12.09                                  & 11.79           & 11.79           & 12.69          & {\color[HTML]{333333} \textbf{13.05}} \\
JPEG2000 \cite{taubman2002jpeg2000}      & 15.57               & 9.18                                   & 8.13            & 8.79            & 8.97           & {\color[HTML]{24292F} 9.57}           \\
WebP \cite{webp}          & 13.92               & 9.09                                   & 8.19            & 8.7             & 9.33           & 9.54                                  \\
BPG \cite{bellard2017bpg}           & 13.26               & 12.98                                & 8.52            & 9.24            & 9.84           & 10.14                                 \\
FLIF \cite{FILF}          & 13.62               & 8.61                                   & 7.44            & 8.16            & 8.73           & 11.7                                  \\
JPEG-XL \cite{alakuijala2019jpeg}        & 14.82               & 8.28                                & 7.2             & 8.19            & 8.49           & $8.61^*$                               \\
JPEG-LS        & 13.35               & -                                      & 7.59            & 8.46            & 8.97           & 9.48                                  \\
LCIC \cite{LCIC}           & -                   & -                                      & 7.88            & 9.02            & 9.35           & -                                     \\
\hline
L3C \cite{l3c}           & 13.26               & 8.97                                   & 7.92            & 8.82            & 9.27           & 9.78                                  \\
SRec \cite{cao2020lossless}           & 12.87               & $8.10^*$                                  & 7.94         & 9.12         & 9.30          &  9.93  \\
RC \cite{RC}           & -                   & -                                      & 7.62            & 8.79            & 9.24           & -                                     \\
IDF \cite{hoogeboom2019integer}           & 11.7                & 8.28                                   &    -             &    -             &                &      -                                 \\
IDF+           & 11.43               &  -                                      &     -            &      -           &   -             &       -                                \\
% Cheng          & 0                   &                                        &                 &                 &                &                                       \\
Near-Lossless \cite{bai2022deepnearlyloss}  & -                  &   -                                     & 7.53            & 7.98            & 8.43           & 9.12                                  \\
iwave++ \cite{iwave}      & 12.5                &                                        &     10.51           &   10.34              &                & 11.87                              \\
% LC-FDNet       & 0-                  &      -                                  & 6.9             & 7.86            & 8.14           & 8.08                               \\
Bit-Swap \cite{Bit-swap}      & 15.18               &      -                                  &     -            &    -             &   -             &   -                                    \\
HiLLoC \cite{townsend2019hilloc}         & 11.7                &     -                                   & -                &   -              &   -             &    -                                 \\
% iwave++(light) & 0                   &                                        &                 &                 &                & 9.21                                  \\

iFlow \cite{iflow}         & $10.95^*$               & -                                      & 6.78            & $7.32$            & 7.71           & -                                     \\
% LLICTI         & -                   & 8.1                                    & -               & -               & -              & -                                     \\
% MSPSM(big)     & -                   & 7.89                                   & -               & -               & -              & -                                     \\

DLPR \cite{bai2024deep}          & 11.07               &  \textbf{7.47} & \textbf{6.48}   & \textbf{7.14}   & $7.65^{*}$  & \textbf{8.58 }                                 \\
\hline
Ours (T=12)          &   \textbf{10.14}                  &   8.19                                    &  6.98              &     7.94           & \textbf{7.62}              & 8.97         \\
\bottomrule
\end{tabular}
}
\end{table*}

\begin{table*}[ht]
\centering
\caption{Average running time (s) of other codecs and ours ($T=12$) on Kodak ($728\times512$) dataset.}
\label{tb:kodak_time}
  % \resizebox{\columnwidth}{!}{ 
\begin{tabular}{cccccccccc}
\toprule
\textbf{runtime(s)} & JPEG-LS & BPG  & FLIF & JPEG-XL & L3C  & Minnen & iWave++ & DLPR & Ours \\
\midrule
Encode              & 0.12    & 2.38 & 0.9  & 0.73    & 8.17  & 2.55   & 158     & 1.26 & 1.12\\
{Decode}     & 0.12    & 0.13 & 0.16 & 0.08    & 7.89    & 5.18   &  -       & 1.8  & 3.54\\
% Params              & -    & - & - & -    & & 5.01M   & 4.20M  & 17.91M    & 1.26 & 1.15 \\
\bottomrule
\end{tabular}
% }
\end{table*}

\begin{figure*}[htbp]  
  \centering
 
\includegraphics[width=0.75\linewidth]{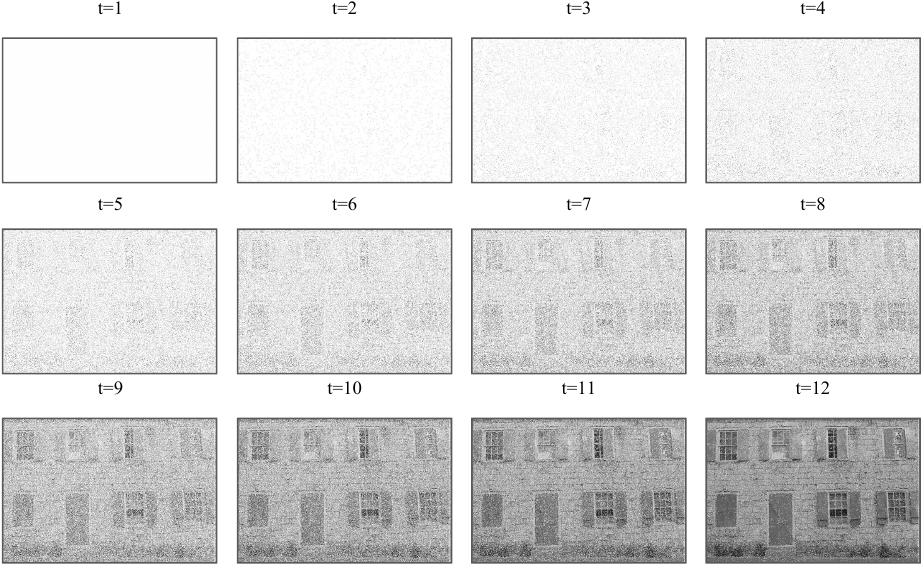}
  \caption{Visualization of coarse-to-fine auto-regressive models. We show the context conditioned on at time $t$, i.e., the encoded/decoded symbols. Unknown symbols are replaced by white pixel dots.}
  \label{fig:corse-to-fine}
\end{figure*}  

\begin{figure}[htbp]
  \centering
\includegraphics[width=0.9\linewidth]{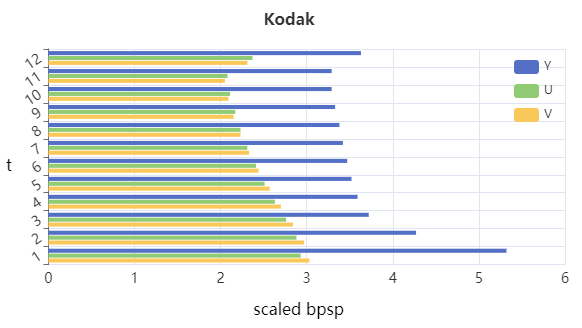}
  \caption{Comparison results (scaled bpsp) for each component and each iteration of the Kodak dataset. The scaled bpsp is the total bits per iteration $t$ divided by the number of coded symbols per iteration $t$ rather than the total number of pixels $N$. }
  \label{fig:yuv_time}
\end{figure}

\begin{figure*}[htbp]
  \centering
\includegraphics[width=0.6\linewidth]{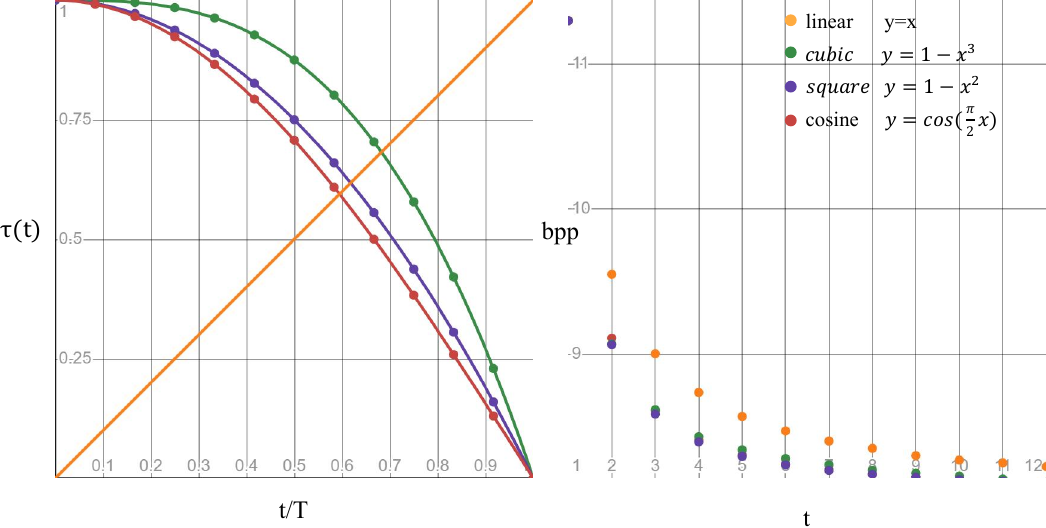}
  \caption{Choices of Mask Scheduling Functions $\tau(t)$ during inference. The right figure shows the compression results (bpp) on the Kodak dataset with different mask functions ($T=12,\beta=10.5$).}
\label{fig:mask_tau}
\end{figure*}

\subsection{Experimental Setup}
\noindent\textbf{Datasets.} 
We conduct a comprehensive evaluation across six benchmark datasets representing diverse image characteristics.
% :We evaluate on six datasets:
\begin{itemize}
\item {\verb|Kodak|}:  Kodak dataset consists of 24 uncompressed $768 \times 512$ color image.

\item {\verb|Open Images|}: The 500 color images selected from the Open Images validation dataset \cite{krasin2017openimages}.
% consists of 24 uncompressed 768×512 color image. dataset consists of 24 uncompressed 768×512 color images. The Open Images Validation Dataset comprises 500 color images, totaling over X million samples, with resolutions ranging from low to ultra-high definition. It serves as a comprehensive benchmark for assessing the performance of computer vision models across a wide range of tasks.
\item {\verb|DIV2K|}: DIV2K high resolution validation dataset \cite{div2k} consists of 100 2K color images. 
% We crop it as 128x128 patches, the same as DPLR \cite{bai2024deep}
\item {\verb|CLIC.p|}: CLIC professional validation dataset \cite{clic} consists of 41 color images, of which most are in 2K resolution.
\item {\verb|CLIC.m|}:  CLIC mobile validation dataset \cite{clic} consists of 61 2K resolution color images.
\item {\verb|ImageNet64|}: ImageNet64 validation dataset is a downsampled variant of the ImageNet validation dataset, consisting of 50000 images of size $64 \times 64$.

\end{itemize}

\noindent\textbf{Training.} 
We train our network on 18000 images chosen from the ImageNet \cite{deng2009imagenet} validation set. During training, we randomly rotate the images and crop patches of the size $256 \times 256$.  The Adam optimizer \cite{kingma2014adam} is used, with the batch-size 16 and initial learning rate $ 1\times10^{-4}$ for 600 epochs. For the sake of stability, we apply gradient clipping (clip max norm =1.0) and decay learning rate to $ 1\times7.5^{-5}$ for the last 300 epochs. We use the ELIC \cite{he2022elic} as our lossy image compressor and freeze it during our training of the residual compressor. Our coding system is trained and evaluated on  Intel(R) Core(TM) i7-9800X CPU @ 3.80GHz, 62G RAM, NVIDIA RTX2080 Ti. We set the same random seed to ensure that each random operation gets replicable results.
% . Note that the mask during training is not generated by the mask generator, but randomly generated, which drives the model to learn various possible situations and improve the generalization.
% The mask during training is randomly generated, which drives the model to learn various possible situations and improve generalization
% \subsubsection{Evaluation} 
% For Bit-Swap, HiLLoC, IDF, IDF++ and LBB, their codes with  numerous parameters can hardly be applied on practical full resolution image compression tasks, and thus only be evaluated on ImageNet64 dataset. For RC, iVPF and iFlow, we report the compression performance published by their authors, because their codes are either difficult to be generalized to arbitrary datasets or unavailable. 
We set $T=12, \beta=10.5$, leading to the best lossless image compression performance.

\subsection{Evaluation Results}
We evaluate the lossless image compression performance
of our coding system with the bit per pixel (bpp). Like DLPR, we also compare with eight traditional lossless image codecs:
including PNG \cite{boutell1997png}, JPEG-LS \cite{jpegls}, CALIC \cite{caclc}, JPEG2000 \cite{taubman2002jpeg2000}, LCIC \cite{LCIC}, WebP \cite{webp}, BPG \cite{bellard2017bpg}, FLIF \cite{FILF} and JPEG-XL \cite{alakuijala2019jpeg}, and 11 recent learning-based lossless image compression methods
including: L3C \cite{l3c}, RC \cite{RC}], Bit-Swap \cite{Bit-swap}, HiLLoC \cite{townsend2019hilloc}, IDF \cite{hoogeboom2019integer}
[14], IDF++ \cite{berg2020idf++}, LBB \cite{LBB}, iFlow \cite{iflow}, iWave \cite{iwave}, DLPR \cite{bai2024deep} and its predecessors Near-Lossless \cite{bai2022deepnearlyloss}. 

As reported in Table~\ref{tab:bpp}, our approach achieves 7.62 bpp, outperforming the previous best DLPR (7.65 bpp) by 0.39\%. This improvement stems from our multidirectional context aggregation that better captures long-range dependencies in high-resolution images. For the ImageNet64 dataset ($64\times64$), we set a new benchmark of 10.14 bpp - an 8.4\% reduction compared to DLPR's 11.07 bpp. This demonstrates the exceptional effectiveness of our method on small images. For other datasets, our method is inferior to DLPR performance but comparable to traditional codecs. The above results demonstrate that our coding system achieves competitive lossless image compression performance and can be effectively generalized to images of various domains.

% validation dataset. T, which shares the same domain with the training dataset. T.shows superior performance to both engineered and learning-based codecs. Considering DIV2K, our method achieves a 3.6\%
% gain compared to Near-Lossless. In the case of CLIC.m, non-learning codecs such as FLIF and JPEG-XL outperform existing learning-based methods. Thus, it can be interpreted that learning-based methods are difficult to be generalized to CLIC.m. Nevertheless, our method achieves state of-the-art performance and outperforms JPEG-XL by 4.3\%. Finally, for CLIC.p, our method shows the best performance
% achieving 1.5\% gain against Near-Lossless.
% In Table 2, we report the compression result for each components for color images Kodak. We observe that the compression efficiency enhances in the order of $Y\to U\to V$. This is straightforward since more information is supplied by the channel context model.
\subsection{Computational Efficiency}
We measure the average runtime required for encoding/decoding the Kodak dataset ($768\times512$). As reported in Table~\ref{tb:kodak_time}, our coding system has superiority over other learned codecs when encoding due to the parallel processing of each component and the pipelined cascade of three operators at each sampling. Since the dependencies between components must be handled serially, our decoding time is longer than DLPR but still comparable to the remaining learning codecs. In detail, for the loss image $\hat{X}$, the average encoding and decoding times are 193 ms and 317 ms. For the residual image $R$, three components can be encoded in parallel, and the total encoding time is 927 ms. The decoding time is 3228 ms due to the serial processing of three components. Together with the lossy image, the total encoding and decoding times are 1120 ms and 3545 ms.

% For the loss image $\hat{X}$, the average encoding and decoding time are 193 ms and 317 ms. At encoding, the components can be operated in parallel, so the total encoding time of 927 ms is taken as the maximum of the 3 components,  $max(\{T*(t^{(c)}_P+t^{(c)}_M)+t^{(c)}_E| c\in [1,2,3]\})$. Since serial decoding of three components, the decoding time of 3228 ms is the sum of the 3 components, $\sum_{c=1}^{3}T*(t_{P}^{c} +t_{M}^{c}+t_{E}^{c})$. Together with the lossy image, the total encoding and decoding times are 1120 ms and 3545 ms, respectively.

\subsection{Coarse-to-fine modeling}
We quantitatively and qualitatively demonstrate the effectiveness of masked sampling in enhancing compression performance. We first report the compression results of each component and iteration in Figure~\ref{fig:yuv_time}. We observe that the compression effects of each component enhance with the increase of t. This is straightforward since more information from more directions becomes available as the iteration proceeds. From the perspective of channels, better compression is presented in the order of $Y \to U\to V$. This can be attributed to our channel context model capturing more dependencies between color components.
Subsequently, we visualize the progressive contexts at each iteration $t$ in Figure~\ref{fig:corse-to-fine}. The reconstructed image becomes clearer over time $t$, which visually demonstrates the learning and enhancement process of our model.

% \subsection
% , which will bring a large increase in running time is not feasible to continue to increase T to reduce BPP, which will bring a large increase in running time Note that the rate of increase in runtime is greater than the rate of decrease in bpp.rate of the runtime grows is greater than the rate at which bpp decreases. Therefore, when T grows to a certain point, the decrease in bpp slows down, and it is not feasible to continue to increase T to reduce the BPP, which would result in a large increase in runtime.

\subsection{Ablation Study}
\noindent\textbf{Different mask schedulers.} We find that the compression bitrate is affected by the mask schedule function $\tau(t)$.
% that computes the masked symbols numbers k. 
As visualized in Figure~\ref{fig:mask_tau}, we consider several functions $\tau(t)$ that satisfy the constraint:  $\tau(t)\in [0,1] and \enspace \tau \downarrow$. The $\tau(t)$ controls the mask ratio in each iteration and influences the $k=\tau(T)=\tau(t)*N$. We observe that the cosine function and square function perform better overall. The linear function performs the worst because it is incapable of scheduling the $k$ well enough to accommodate our coarse-to-fine process.

\noindent \textbf{Rate-latency tradeoff Analysis .}
We show the results of adjusting variable $\textit{T}$ to trade off bitrate and coding runtime in Table~\ref{bpp_t}. As $T$ increases, the bpp decreases, and the coding runtime increases. When T increases to a certain extent, the reduction of bpp slows down and the growth of runtime accelerates.  Therefore, we cannot continue to increase  $T$ to reduce bpp at this point, which will bring a large increase in running time. All things considered, $T=12$ is a good balance between rate and runtime.
% \begin{table}[]
% \caption{The effects of different mask schedule function $\tau(t)$  and temperature coefficient $\beta$ on lossless image compression performance
% (bpp), which only includes the residual part.}
% \begin{tabular}{lllllll}
% \toprule
% \textbf{$\tau(t)$}   &$\beta=1.5$ & $\beta=4.5$    &$\beta=7.5$  & $\beta=10.5$   & $\beta=12.5$  & $\beta=5.5$   \\
% \midrule
% Linear          &       & 8.23       &       &        &        &        \\
% Square&8.17       & 8.12  & 8.14 & 8.11 & /       &8.11        \\
% Cubic           &  8.24     & 8.16     & 8.14     & 8.13 & 8.11       & 8.11       \\
% Cosine          & 8.16 & 8.12 & 8.11 & 8.11 & 8.11 & 8.11\\
% \bottomrule
% \end{tabular}
% \end{table}

\begin{table}[ht]
% \label{tab:dirrentT}
\caption{The effects of different $\textit{T}$ for bpp and runtime (s) on Kodak with cosine mask schedule function and temperature coefficient $\beta=10.5$ (only including the residual part of image).}
\begin{tabular}{ccccccc}
\toprule
% \textbf{$T=12$}   & \multicolumn{6}{c}{$\beta$}                   \\
% \midrule
% $\tau(t)$           & 1.5  & 4.5  & 7.5  & 10.5 & 12.5 & 15.5 \\
%  \cmidrule(lr){2-7}
% Linear & 8.26 & 8.23 &   8.22   &   8.23   & 8.23     &8.23      \\
% Square          & 8.17 & 8.12 & 8.14 & 8.11 &   8.11   & 8.11 \\
% Cubic           & 8.24 & 8.16 & 8.14 & 8.14 & 8.11 & 8.11 \\
% Cosine          & 8.16 & 8.12 & 8.11 & 8.11 & 8.11 & 8.11 \\
% \hline
% \makecell[c]{Cosine\\$\beta=10.5$ }  & \multicolumn{6}{c}{$T$}                   \\
% \hline
                & 1    & 2    & 5    & 8    & 12   & 32   \\
                \midrule
                
bpp             &  11.30    & 9.11     &  8.30    &  8.17    & 8.11     &    8.04  \\
Encode (s)       &  0.272    &  0.380    & 0.458     & 0.733    & 0.933     &   2.451   \\
Decode (s)       & 0.760     &1.236      & 1.368     &  2.189    &   3.246   & 8.758    \\
\bottomrule
\end{tabular}
\label{bpp_t}
\end{table}

% \begin{table}[ht]
% \label{tab:dirrentT}
% \caption{The effects of different mask schedule function $\tau(t)$  and temperature coefficient $\beta$ on lossless image compression performance
% (bpp), which only includes the residual part.}
% \begin{tabular}{ccccccc}
% \toprule
% \textbf{$T=12$}   & \multicolumn{6}{c}{$\beta$}                   \\
% \midrule
% $\tau(t)$           & 1.5  & 4.5  & 7.5  & 10.5 & 12.5 & 15.5 \\
%  \cmidrule(lr){2-7}
% Linear & 8.26 & 8.23 &   8.22   &   8.23   & 8.23     &8.23      \\
% Square          & 8.17 & 8.12 & 8.14 & 8.11 &   8.11   & 8.11 \\
% Cubic           & 8.24 & 8.16 & 8.14 & 8.14 & 8.11 & 8.11 \\
% Cosine          & 8.16 & 8.12 & 8.11 & 8.11 & 8.11 & 8.11 \\
% \hline
% \makecell[c]{Cosine\\$\beta=10.5$ }  & \multicolumn{6}{c}{$T$}                   \\
% \hline
%                 & 1    & 2    & 5    & 8    & 12   & 32   \\
%                 \cmidrule(lr){2-7}
% bpp             &  11.30    & 9.11     &  8.30    &  8.17    & 8.11     &    8.04  \\
% Encode(s)       &  0.272    &  0.380    & 0.458     & 0.733    & 0.933     &   2.451   \\
% Decode(s)       & 0.760     &1.236      & 1.368     &  2.189    &   3.246   & 8.758    \\
% \bottomrule
% \end{tabular}
% \end{table}
\section{Conclusion}
In this paper, we have proposed a scalable lossless image compression framework that supports single/multiple-component images with a single-trained model. Our framework consists of a lossy image compressor and a specialized residual compressor, which decomposes the whole residual compression into  $T$ times masked sampling. With the mask design, we propose the multiple-direction auto-regressive context model for probability estimation of residual.
The  $T$ times masked sampling proceeds in a coarse-to-fine manner, which progressively refines the probability estimation of residual. Furthermore, our framework enables tunable rate-latency tradeoffs through adaptive iteration times $T$.
% Moreover, we resolve the defect of the existing methods only considering the one-direction context in scan order by the proposed multiple-direction auto-regressive context model.  
Experiments demonstrate
that our coding framework implements competitive lossless compression performance with practical complexity on various full-resolution images.
% of variable rate to adapt to compression requirements of different scenarios. Experiments show
% that our method achieves superior performance for extensive datasets.
\subsection{Discussion}
Due to the training cost, we freeze the lossy compression network and only train the proposed residual network. If these models are trained jointly, the compression performance will be further improved. In addition, how to train the residual compressor generalized to residuals of different distributions produced by various lossy compression networks, which is also the direction of our efforts.

% \section{Acknowledgments}

% \begin{acks}
% \end{acks}

\bibliographystyle{ACM-Reference-Format}
\bibliography{main}

\end{document}